\documentclass{sig-alternate}
\usepackage{times,amsmath,epsfig}
\usepackage{latexsym}
\usepackage{graphicx}
\usepackage{amsfonts}
\usepackage{amssymb}

\usepackage{subfigure}
\usepackage{algorithm}
\usepackage[noend]{algorithmic}
\usepackage{xspace}

\usepackage{multirow}

\begin{document}

\title{Context-based Diversification for Keyword Queries over XML Data}
\numberofauthors{4} 

\author{%
\alignauthor Jianxin Li\\
					\affaddr{Swinburne University of Technology}\\
					\affaddr{Melbourne, Australia}\\
					\email{jianxinli, cliu, liangyao@swin.edu.au}
\alignauthor Chengfei Liu \\
					\affaddr{Swinburne University of Technology}\\
					\affaddr{Melbourne, Australia}\\
					\email{cliu@swin.edu.au}
\alignauthor Liang Yao \\
					\affaddr{Swinburne University of Technology}\\
					\affaddr{Melbourne, Australia}\\
					\email{liangyao@swin.edu.au}
\and 
\alignauthor Jeffrey Xu Yu \\
					\affaddr{Chinese University of Hong Kong}\\
					\affaddr{Hong Kong, China}\\
					\email{yu@se.cuhk.edu.hk}
}

\date{18-23 March 2013}

\newtheorem{definition}{Definition}
\newtheorem{theorem}{Theorem}
\newtheorem{example}{Example}
\newtheorem{corollary}{Corollary}
\newtheorem{lemma}{Lemma}
\newtheorem{property}{Property}

\maketitle

\begin{abstract}

While keyword query empowers ordinary users to search vast amount of data,
the ambiguity of keyword query makes it difficult to effectively answer keyword queries, especially for short and vague keyword queries. To address this challenging problem, in this paper we propose an approach that automatically diversifies XML keyword search based on its different contexts in the XML data. Given a short and vague keyword query and XML data to be searched, we firstly derive keyword search candidates of the query by a classific feature selection model. And then, we design an effective XML keyword search diversification model to measure the quality of each candidate. After that, three efficient algorithms are proposed to evaluate the possible generated query candidates representing the diversified search intentions, from which we can find and return top-$k$ qualified query candidates that are most relevant to the given keyword query while they can cover maximal number of distinct results.
At last, a comprehensive evaluation on real and synthetic datasets demonstrates the effectiveness of our proposed diversification model and the
efficiency of our algorithms.
\end{abstract}


\section{Introduction}



Keyword search on structured and semi-structured data has attracted much research interest recently, as it enables common users to retrieve information from such structured data sources without the need to learn sophisticated query languages and database structure~\cite{DBLP:conf/sigmod/ChenWLL09}. 
In general, the more keywords a given keyword query contains, the easier the search semantics of the keyword query can be identified. 
However, when the given keyword query only contains a small number of vague keywords, it will become a very challenging problem to derive the search semantics of the query due to the high ambiguity of this type of keyword queries.
Although sometimes user involvement is helpful to identify search semantics of keyword queries, it is not always applicable to rely on users because 
the keyword queries may also come from system application. 
In this application case, web or database search engine may need to automatically compute the search semantics of short and frequent keyword queries only based on the data to be searched. The derived search semantics will be maintained and updated in an off-line way. Once a keyword query is issued by the real users, its corresponding search semantics can be directly used to make an instant response. In this paper, we mainly pay attention to the problem of effectively deriving the search semantics of keyword queries with the consideration of data only, which does not receive much closer attention in the previous works.

\begin{example}\label{exam:motivation}
Consider a simple keyword query $q$=\{database, query\} over the DBLP dataset. There are 21,260 publications containing the keyword ``database'', and 9,896 publications containing the keyword ``query'', which contributes 2040 results that contain the two given keywords together. If we directly explore and understand the keyword search results, it would be time consuming and not user-friendly due to the huge number of results. It needs to take 54.22 seconds for just computing all the SLCA results of $q$ by using XRank~\cite{DBLP:conf/sigmod/GuoSBS03}. Even if the system processing time is acceptable by accelerating the keyword query evaluation with efficient algorithms~\cite{DBLP:conf/www/SunCG07,DBLP:conf/sigmod/XuP05}, the unclear and repeated search intentions in the large set of retrieved results will make users frustrating. To address the problem,  
we derive different search semantics of the original query from the different contexts of the XML data to be searched, which can be used to represent the different search intentions of the original query. In this work, the contexts can be modeled by extracting some relevant feature terms of the query keywords from the XML data, as shown in Table~\ref{tab:features}. And then, we compute the keyword search results for each search intention. Table~\ref{fig:snapshot} shows part of statistic information of the answers related to the keyword query $q$, which classifies each ambiguous keyword query into different search intentions. 
\end{example}

\begin{table}[t]
  \renewcommand{\arraystretch}{1.2}
  \small
  \centering
  \caption{Top 10 selected feature terms of $q$}
  \label{tab:features}
    \scalebox{1}{
  \begin{tabular}{|l|l|}  
    \hline
    \textbf{keyword} & \textbf{features} \\[0.5ex]
    \hline
    database & \textit{systems}; \textit{relational}; \textit{protein}; \textit{distributed}; \\ 
    & \textit{oriented}; \textit{image}; \textit{sequence}; \textit{search}; \\
    & \textit{model}; \textit{large}. \\ 
    \hline
    query &   \textit{language}; \textit{expansion}; \textit{optimization}; \textit{evaluation}; \\
    & \textit{complexity}; \textit{log}; \textit{efficient}; \textit{distributed}; \\ 
    & \textit{semantic}; \textit{translation}. \\ 
   
    \hline
     \end{tabular}
      }
\end{table}

\begin{table}[t]
  \renewcommand{\arraystretch}{1.2}
  \small
  \centering
  \caption{Part of statistic information for $q$}
  \label{fig:snapshot}
  \scalebox{0.85}{
  \begin{tabular}{|l|ccccc|}  
    \hline
     					& \multicolumn{5}{|c|}{database \textit{systems} query + } \\
     					\cline{2-6}
		 & \textit{language} & \textit{expansion}& \textit{optimization}& \textit{evaluation}&\textit{complexity} \\
	   \#\textbf{results}   & 71 & 5& 68 & 13 & 1 \\
		\hline
		&\textit{log}& \textit{efficient} & \textit{distributed}& \textit{semantic} & \textit{translation}\\
			\#\textbf{results}  &12&17&50&14&8\\		
		\hline
		& \multicolumn{5}{|c|}{\textit{relational} database query + } \\
			\cline{2-6}
		& \textit{language} & \textit{expansion}& \textit{optimization}& \textit{evaluation}&\textit{complexity} \\
		\#\textbf{results}  & 40 & 0& 20 & 8 & 0 \\
		\hline
		&\textit{log}& \textit{efficient} & \textit{distributed}& \textit{semantic} & \textit{translation}\\
		 \#\textbf{results}   &2&11&5&7&5\\   
    \hline
    ... &  ... & &&&\\
    \hline
  \end{tabular}
  }
\end{table}

By exploring the different feature terms of the query keywords, we have two benefits: the first is to diversify the keyword search results automatically by the different search intentions, which can return more distinct and diversified results to users; and the second is to improve the efficiency of keyword search because the contexts of diversified keyword queries can be used to reduce the size of relevant keyword node lists.  

%

Therefore, we are motivated to study the problem of keyword search diversification based on the contexts of query keywords in XML data to be searched, which is denoted as \textit{intent-based diversification}.
Although the
intent-based diversification has been discussed in information retrieval (IR), e.g.,  \cite{DBLP:conf/wsdm/AgrawalGHI09} models user intents at the topical level of the taxonomy and \cite{DBLP:conf/sigir/RadlinskiD06} obtains the possible query intents by mining query
logs, they are not always applicable because on the one hand, it is not easy to get the useful taxonomy and query logs; on the other hand, diversified results are modelled at different level, i.e., documents in IR vs. fragments in XML. 
To the best of our knowledge, \cite{DBLP:conf/sigir/DemidovaFZN10} is the most relevant work that first maps each keyword to a set of attribute-keyword pairs, and then constructs a set of structured queries. It assumes that each structured query represents a query interpretation. However, the assumption is too strict to be applied for XML data because contextual information may not be necessarily structured, i.e., it may appear in the form of either attribute labels or texts. 

The problem of diversifying keyword search is firstly proposed and studied in IR community \cite{DBLP:conf/sigir/CarbonellG98,DBLP:conf/wsdm/AgrawalGHI09,DBLP:conf/sigir/ChenK06,DBLP:conf/sigir/ClarkeKCVABM08,DBLP:conf/sigmod/AngelK11}.
Most of the techniques perform diversification as a post-processing or re-ranking step of document retrieval based on the analysis of result set and/or the historic query logs.
In IR, keyword search diversification is designed at the topic or document level.
For structured databases or semi structured databases, it is necessary to be redesigned at the tuple or fragment level. 
To address the main difference, the authors in \cite{DBLP:conf/sigmod/ChenL07} propose to navigate SQL
results through categorization, which takes into account user preferences. It consists of two steps: the first step analyzes query history of all users
in the system offline and generates a set of clusters over the data, each
corresponding to one type of user preferences; for an issued query, the 
second step presents to the user a navigational tree over clusters generated
in the first step. By doing this, the user can browse, rank, or categorize 
the results in selected clusters.
The authors in \cite{DBLP:conf/icde/VeeSSBA08} introduce a pre-indexing 
approach for efficient diversification of query results on relational 
databases based on the prespecified diversity orderings among the attributes over relations.
The authors in \cite{DBLP:journals/pvldb/LiuJ09} first work out a small number of tuples by choosing one
representative from each of clusters and return them in the first page, 
which helps users learn what is available in the whole result set and 
directs them to find what they need. The authors in \cite{DBLP:journals/pvldb/LiuSC09} differentiate the keyword search results by comparing their feature sets where their feature types are limited to the labels of XML elements in the keyword search results.
All of these methods can be classified as \textit{post-process search result analysis}. They will encounter two challenging problems: the first one is effectiveness because the comparison of results will become difficult when the content of a result is not too much informative; the second is efficiency because they have to compute all the results, analyse and compare them one by one.



To address the above limitations, we initiate a formal study of the diversification problem in XML keyword search, which can directly compute the diversified results without retrieving all the relevant candidates. Towards this goal, given a keyword query, we
first derive the co-related feature terms for each query keyword from the XML data based on the mutual information in probability theory, which has been used as a criterion for feature selection \cite{DBLP:journals/pami/PengLD05,DBLP:conf/icpr/SakarK10}.
The selection of our feature terms is not limited to the labels of XML elements.
Each combination of the feature terms and the original query keywords represents one of diversified contexts that express specific search intentions.
And then, we evaluate derived search intentions by considering their relevances to the original keyword query and the novelty of the produced results.
To efficiently compute diversified keyword search, we propose one baseline algorithm and two efficient algorithms based on the observed properties of diversified keyword search results.

The remainder of this paper is organized as follows. In Section~\ref{sec:problemdefinition}, we introduce a feature selection model and define the problem of diversifying XML keyword search. 
We describe the procedure of extracting the relevant feature terms for a keyword query based on the explored feature selection model in Section~\ref{sec:infersurprise}.
 In Section~\ref{sec:ksdalgorithm}, we first show the procedure of generating search intentions from the derived feature terms and then propose three efficient algorithms, based on the observed properties of XML keyword search results, to identify a set of qualified and diversified keyword queries and compute their corresponding results. In Section~\ref{sec:experiments}, we provide extensive experimental results to show the effectiveness of our XML keyword search diversification model and the performance of our proposed algorithms. We describe the related work in Section~\ref{sec:relatedwork} and conclude in Section~\ref{sec:conclusion}.   
  


\section{Problem Definition}\label{sec:problemdefinition}
Given a keyword query $q$ and an XML data denoted by $T$, we consider a set of possible search intentions $Q$ that are generated by bounding each query keyword to a context using its relevant feature terms in $T$. 
Here, search intentions are also represented in the format of keyword query. Naturally, we need present to the users the top $k$ qualified queries in terms of high relevance and maximal diversification.    
\subsection{Feature Selection Model}
Consider an XML data $T$ and a set of term-pairs 
$W$ that 
can appear in $T$. The composition method of $W$ depends
on the application context and will not affect our subsequent discussion. As
an example, it can simply be the full or a subset of the terms comprising the text in $T$, the contents of a dictionary, or a well-specified set of 
term-pairs
relevant to some applications. 



In this work, the distinct term-pairs are selected 
based on their mutual information as 
~\cite{DBLP:journals/pami/PengLD05,DBLP:conf/icpr/SakarK10}. Mutual information has been used as a criterion for feature selection and feature transformations in machine learning. It can be used to characterize both the relevance and redundancy of variables, such as the minimum redundancy feature selection.
Assume we have an XML tree $T$ and its sample result set $R(T)$.
Let $Prob(x, T)$ be the probability of term $x$ appearing in $R(T)$, i.e., $Prob(x, T) = \frac{|R(x, T)|}{|R(T)|}$ where $|R(x, T)|$ is the number of results containing $x$. 
Let $Prob(x, y, T)$ be the probability of terms $x$ and $y$ co-occurring in $R(T)$, i.e., $Prob(x, y, T) = \frac{|R(x, y, T)|}{|R(T)|}$. If terms $x$ and $y$ are independent, then knowing $x$ does not give any information about $y$ and vice versa, so their mutual information is zero. At the other extreme, if terms $x$ and $y$ are identical, then knowing $x$ determines the value of $y$ and vice versa. Therefore, the simple measure can be
used to quantify by how much the observed word co-occurrences that maximize the dependency of feature terms while reduce the redundancy of feature terms.
In this work, we use the popularly-accepted mutual information model as follows.

\begin{equation}\label{equ:surprise}
\begin{array}{l} 
MI(x, y, T) = Prob(x, y, T)* log\frac{Prob(x, y, T)}{Prob(x, T)*Prob(y, T)}  
\end{array}
\end{equation}

For each term in the XML data, we need to find a set of feature terms where the feature terms can be selected in any way, e.g., top-$m$ feature terms or the feature terms with their mutual values higher than a given value based on domain applications or data administrators. The feature terms can be pre-computed and stored before the procedure of query evaluation. Thus, given a keyword query, we can obtain a matrix of features for the query keywords against the XML data to be searched.
The matrix constructs a space of search intentions of the original query w.r.t. the XML data. Therefore, our first problem is to find a set of feature terms from the matrix, which has the highest probability of interpreting the contexts of original query. In this work, we extract and evaluate the feature terms at the entity level of XML data.

\begin{table}[thbc]
  \renewcommand{\arraystretch}{1.2}
  \small
  \centering
  \caption{Mutual information score w.r.t. terms in $q$}
  \label{tab:mutualscore}
  \scalebox{0.7}{
  \begin{tabular}{|l|ccccc|}  
    \hline
    \multirow{2}{*}{\textbf{database}} &  \textit{system} & \textit{relational}& \textit{protein}& \textit{distributed}&\textit{oriented} \\
  															& 7.06  & 3.84 & 2.79 & 2.25 & 2.06 \\
    \cline{2-6}
   \multirow{2}{*}{\textbf{Mutual score} ($10^{-4}$)} & \textit{image}& \textit{sequence} & \textit{search}& \textit{model} & \textit{large}\\
    
		 														 &1.73&1.31&1.1&1.04&1.02\\
		 \hline
		\multirow{2}{*}{\textbf{query}} & \textit{language} & \textit{expansion}& \textit{optimization}& \textit{evaluation}&\textit{complexity}\\
		 											  & 3.63 & 2.97& 2.3 & 1.71 & 1.41 \\
		 											  \cline{2-6}
 \multirow{2}{*}{\textbf{Mutual score} ($10^{-4}$)}& \textit{log}& \textit{efficient} & \textit{distributed}& \textit{semantic} & \textit{translation}\\
									&1.17&1.03&0.99&0.86&0.70\\	   
    \hline
  \end{tabular}}
\end{table}

Consider query $q$ =\{database, query\} over DBLP XML dataset again. Its corresponding matrix can be constructed from Table~\ref{tab:features}. Table~\ref{tab:mutualscore} shows the mutual information score for the query keywords in $q$. Each combination of the feature terms in matrix represents a search intention with the specific semantics. 
For example, the combination ``query \textit{expansion} database \textit{systems} '' targets to search the publications discussing the problem of query expansion in the area of database systems, e.g., one of the works, ``\textit{query expansion for information retrieval}'' published in \textit{Encyclopedia of Database Systems} in 2009, will be returned. If we replace the feature term ``systems'' with ``relational'', then the generated query will be changed to search specific publications of query expansion over relational database, in which the returned results are empty because no work is reported to the problem over relational database in DBLP dataset.


\subsection{Keyword Search Diversification Model}

In this model, we not only consider the probability of new generated queries, i.e., \textit{relevance}, we also take into account their new and distinct results, i.e., \textit{novelty}.
To embody the relevance and novelty of keyword search together, two criteria should be satisfied: (1) the generated query $q_{new}$ 
has the maximal probability to interpret the contexts of original query $q$ with 
regards to the data to be searched; and (2) the generated 
query $q_{new}$ has a maximal difference from the previously generated query set $Q$.
 Therefore, we have the aggregated scoring function.
 

\begin{equation}\label{eq:nextquery}
\begin{array}{l}
score(q_{new}) = Prob(q_{new}|q,T) * DIF(q_{new},Q,T)
\end{array}
\end{equation}
where $Prob(q_{new}|q,T)$ represents the probability that $q_{new}$ is the 
search intention when the original query $q$ is issued over the data $T$; $DIF(q_{new},Q,T)$ represents the percentage of results that are produced by $q_{new}$, but not by any generated query in $Q$.




Firstly, let's show how to calculate the probability $Prob($ $q_{new} | q$, $T)$
 of a query $q_{new}$ that is intended while the user issues $q$ on the XML data $T$. 
 Based on the Bayes Theorem, we have

\begin{equation}\label{equ:generatedprob}
\begin{array}{l}
Prob(q_{new}|q,T) = \frac{Prob(q|q_{new}, T) * Prob(q_{new}|T)}{Prob(q|T)}
\end{array}
\end{equation}
where $Prob(q|q_{new}, T)$ models the likelihood of generating the observed query $q$ while the intended query is actually $q_{new}$; and $Prob(q_{new}|T)$ is the query generation probability given the XML data $T$.


The likelihood value $Prob(q|q_{new}, T)$ can be measured by computing the probability of the original query $q$ that is observed in the context of the features in $q_{new}$.  
Given a query $q=\{k_i\}$ and a generated new query $q_{new}=\{s_i\}$ where $k_i$ is a query keyword in $q$, $s_i$ is a segment that consists of the query keyword $k_i$ and one of its features $f_{ij_i}\in P$, $1\leq i \leq n$, and $1\leq j_i \leq m$. Here, we assume that for each query keyword, only top $m$ most relevant features will be retrieved from $P$ to generate new queries. 
To deal with multi-keyword queries, we make the independence assumption on the probability of generating a query keyword $k_i$ while the intended feature is actually $f_{ij_i}$. That is,
\begin{equation}\label{equ:queryprob}
\begin{array}{l}
Prob(q|q_{new}, T) = \prod_{k_i\in q, f_{ij_i}\in q_{new}} Prob(k_i|f_{ij_i}, T)
\end{array}
\end{equation}

According to the statistical information, the intent of a keyword can be 
inferred from the occurrences of the keyword and its correlated terms in 
the data to be searched. Thus, we can compute the probability $Prob(k_i|f_{ij_i}, T)$ of interpreting a keyword $k_i$ into a search intent $f_{ij_i}$ as follows.
\begin{equation}\label{equ:keywordintent} 
\begin{array}{rl}
Prob(k_i|f_{ij_i}, T) = & \frac{Prob(f_{ij_i}|k_i, T)*Prob(k_i, T)}{Prob(f_{ij_i}, T)} \\
= &\frac{|R(\{k_i,f_{ij_i}\}, T)|/|R(T)|}{|R(f_{ij_i}, T)|/|R(T)|}\\
 = &\frac{|R(\{s_i\}, T)|}{|R(f_{ij_i}, T)|}
\end{array}
\end{equation}
where $s_i = \{k_i,f_{ij_i}\}$.

Consider a query $q$ =\{database, query\} and one of its new queries $q_{new}$=\{database system; query expansion\}. $Prob($ q$|q_{new},$ $T)$ show the probability of a publication that addresses the problem of ``database query'' regarding the context of ``system and expansion'', which can be computed by $\frac{|R(\{\text{database system}\}, T)|}{|R(\text{system}, T)|}$ * $\frac{|R(\{\text{query expansion}\}, T)|}{|R(\text{expansion}, T)|}$. 
Here, $|R(\{\text{database system}\}, T)|$ represents the number of keyword search results  of query \textit{database system}\} over the data $T$. The result type can be defined by users. In this work, we adopt the widely accepted semantics - \textit{Smallest Lowest Common Ancestor} {SLCA} to model XML keyword search results. Briefly, a node $v$ is regarded as an SLCA for a keyword query if (a) the subtree (T$_{sub}$($v$)) rooted at the node $v$ contains all the query keywords; (b) there does not exist a descendant node $v'$ of $v$ such that T$_{sub}$($v'$) contains all the query keywords.
$|R(\text{system}, T)|$ represents the number of keyword search results of running \textit{system} over the data $T$, but the number can be obtained without running query \textit{system} because it is equal to the size of keyword node list of ``system'' over $T$.

Given the XML data $T$, the query generation probability of $q_{new}$ can be calculated by. 

\begin{equation} 
\begin{array}{l}
Prob(q_{new}|T) = \frac{|R(q_{new}, T)|}{|R(T)|} = \frac{|\bigcap_{s_i\in q_{new}}R(s_i, T)|}{|R(T)|}
\end{array}
\end{equation}
where $\bigcap_{s_i\in q_{new}}R(s_i, T)$ represents the set of SLCA results by merging the node lists $R(s_i, T)$ for $s_i\in q_{new}$ using the algorithms~\cite{DBLP:conf/www/SunCG07,DBLP:conf/sigmod/XuP05}. 

Given an original query and the data, the value $\frac{1}{Prob(q|T)}$ is a relatively unchanged value with regards to different generated queries. Therefore, the above equation can be rewritten as follows.
\begin{equation}\label{equ:probq}
\begin{array}{l}
Prob(q_{new}|q,T) =  \gamma*(\prod\frac{R(s_i, T)}{R(f_{ij_i}, T)}) *  \frac{|\bigcap R(s_i, T)|}{|R(T)|} 
\end{array}
\end{equation}
where $k_i\in q$, $s_i\in q_{new}$, $f_{ij_i}\in s_i$ and $ \gamma=\frac{1}{Prob(q|T)}$ can be any value in (0,1] because it does not affect the refined query selection w.r.t. an original keyword query $q$ and data $T$. 

  
Although the above equation can model the probability of possible intended query, i.e., the relevance between the generated queries and the original query w.r.t. the data, different generated queries may return overlapped result sets. Therefore, we should also take into account the novelty of results produced from the generated queries. 


Now let's show how to quantify the value $DIF(q_{new},Q,T)$. Given a single keyword query, it can be answered based on the normal SLCA semantics. However, the normal SLCA semantics is not enough to explain the results in this paper because by the SLCA result of a query $q$, we mean the SLCA results of the set of all generated queries of $q$. Let $v$ be an SLCA node returned from the previously generated query $q_{pre}$ of $q$, and $v'$ be an SLCA node returned from the newly generated query $q_{new}$ of $q$. Similar to the normal SLCA semantics, if $v'$ is an ancestor node of $v$ or the same as $v$, $v'$ cannot become a new SLCA node of $q$. However, different from the normal SLCA semantics, if $v'$ is a descendant node of $v$, $v'$ will replace $v$ as a new SLCA node of $q$ because $v'$ presents more specific semantics that $v$ does. 
As such, the novelty $DIF(q_{new},Q,T)$ of a new generated query $q_{new}$ against the previously generated query set $Q$ can be calculated as follows.

\begin{equation}\label{equ:approximatedif}
\begin{array}{l}
DIF(q_{new}, Q, T) = 
\\
 \frac{|\{v_x| v_x\in R(q_{new}, T) \wedge \nexists v_y\in \{\bigcup_{q'\in Q}R(q', T)\} \wedge v_x\leq v_y\}|}{|R(q_{new}, T)\bigcup \{\bigcup_{q'\in Q}R(q', T)\}|}
\end{array}
\end{equation}
where $v_x\leq v_y$ means that $v_x$ is a duplicate of $v_y$, i.e., ``='', or $v_x$ is an ancestor of $v_y$, i.e., $``<''$; $\bigcup_{q'\in Q}R(q', T)$ represents the set of SLCA results generated by queries in $Q$, which have to clean up duplicate SLCA nodes and ancestor SLCA nodes in the SLCA result set of $Q$.





As we know our problem is to find top $k$ qualified queries and their relevant SLCA results. To do this, we can compute the absolute score of the search intention for each generated query. To reduce the computational cost, an alternative way is to calculate the relative scores of queries. Therefore, we have the following equation transformation. After we substitute Equation~\ref{equ:probq} and Equation~\ref{equ:approximatedif} into Equation~\ref{eq:nextquery}, we have the final equation.
\begin{equation}\label{eq:nextquery2}
\begin{array}{l}
score(q_{new})=  \gamma *\prod(\frac{R(s_i, T)}{R(f_{ij_i}, T)}) *  \frac{|\bigcap R(s_i, T)|}{|R(T)|} * \\

   \frac{|\{v_x| v_x\in R(q_{new}, T) \wedge \nexists v_y\in \{\bigcup_{q'\in Q}R(q', T)\} \wedge v_x\leq v_y\}|}{|R(q_{new}, T)\bigcup \{\bigcup_{q'\in Q}R(q', T)\}|} \\
   
=  \frac{\gamma}{|R(T)|} *\prod(\frac{R(s_i, T)}{R(f_{ij_i}, T)}) *  |\bigcap R(s_i, T)| * \\

   \frac{|\{v_x| v_x\in R(q_{new}, T) \wedge \nexists v_y\in \{\bigcup_{q'\in Q}R(q', T)\} \wedge v_x\leq v_y\}|}{|R(q_{new}, T)\bigcup \{\bigcup_{q'\in Q}R(q', T)\}|} \\
 
 \mapsto  \prod(\frac{R(s_i, T)}{R(f_{ij_i}, T)}) *  |\bigcap R(s_i, T)| * \\
   \frac{|\{v_x| v_x\in R(q_{new}, T) \wedge \nexists v_y\in \{\bigcup_{q'\in Q}R(q', T)\} \wedge v_x\leq v_y\}|}{|R(q_{new}, T)\bigcup \{\bigcup_{q'\in Q}R(q', T)\}|}
\end{array}
\end{equation}
where $k_i \in q$, $s_i \in q_{new}$, $f_{ij_i}\in s_i$, $q' \in Q$ and the symbol $\mapsto$ represents the left side of the equation \textit{depends on} the right side of the equation because the value $\frac{\gamma}{|R(T)|}$ keeps unchanged for calculating the diversification scores of different search intentions.
\section{Extracting Feature Terms}\label{sec:infersurprise}

Although we can pre-compute and manipulate the co-related terms up to any size, the use of two-term co-occurrences presents the most reasonable
alternative in most applications \cite{DBLP:journals/pvldb/SarkasBDK09}. In 
addition, two-term co-occurrences can be computed and stored efficiently as described in \cite{DBLP:conf/vldb/BansalCKT07}. Co-occurrences of higher order can be 
utilized at the expense of space and, most importantly, time. For the scale
of the applications we envision, materializing co-occurrences of length higher than two is probably infeasible. Therefore, in this work, we materialize two-term co-occurrences, which involves the computation of a 
sorted list consisting of triplets ($x$, $y$, $R(x,y)$). Every
such triplet contains the set of IDs of the results for terms $x$ and $y$. The triplets are sorted based on the result size. If two terms do not co-occur, we simply don't store the corresponding tuple.

To infer the feature terms for a keyword, we first take the entity nodes (e.g., the nodes with the ``*'' node types in XML DTD) as a sample space. The inferred terms represent different feature interpretations of the keyword in the XML data to be searched. For every possible keyword, it has a list of entity IDs where the entity ID is encoded using Dewey Coding Scheme. Given a keyword $x$ and a term $y$, their mutual information score can be calculated based on 
Equation~\ref{equ:surprise} where $Prob(x, T)$ (or $Prob(y, T)$) is the value of dividing the list size of $x$ (or $y$) by the total entity size of the sample space;
$Prob(\{x,y\}, T)$ is the value of dividing the merged result size of lists $x$, $y$ by the total entity size of the sample space.
Similarly, we can compute the mutual information scores of $x$ with the other terms. After that, the terms with the top-$m$ mutual information scores will be selected as the $m$ distinct features of the keyword $x$ in the sample space.

To efficiently compute the mutual information score, we only need to traverse the XML data tree once. During the XML data tree traversal, we first extract the meaningful text information from the entity we met. Here, we would like to filter out the stop words, which cannot contribute to extract meaningful contexts of search intentions. And then we produce a set of term-pairs by scanning the extracted text. 
To maximize the relevance and reduce the redundancy, we set a limitation that the position of a term in the text has maximal three 
 steps to another term that can cooccur in the same term-pair. The maximal steps can be varied based on requirements.
After that, all the reasonable term-pairs will be generated and recorded. When the next entity comes, we will process its text in the same way and generate meaningful term-pairs. If a term-pair has been recorded beforehand, then we just increase the count of the term-pair. Otherwise, a new term-pair will be created and recorded. At the same time, the total number of entities will be recorded, too. After the XML data tree is traversed completely, we can compute the mutual information score for each term-pair based on Equation~\ref{equ:surprise}.

  
  
     

\section{Keyword Search Diversification Algorithms}\label{sec:ksdalgorithm}

In this section, we first introduce the procedure of generating a new query 
from the matrix of the original keyword query w.r.t. the data to be searched. And then based on the matrix, we propose a baseline algorithm to retrieve the diversified keyword search results. At last, two anchor-based pruning algorithms are designed to improve the efficiency of the keyword search diversification by utilizing the intermediate results.

\subsection{Generate Search Intentions}

Given a keyword query $q$, we first retrieve the corresponding feature terms for each query keyword and then construct a matrix of search intentions. In the matrix, the feature terms in each
column are sorted based on their mutual information scores. Each combination of the feature terms (one term per column) represents a search intention. We 
iteratively choose the combination with the maximal aggregated mutual information score as the next best search intention until the terminal requirements are reached.

As we discussed above, the aggregated mutual information score of each search intention represents to some extent the confidence of the context of the query keywords. Without other knowledge, we would like to generate the search intentions and then check the corresponding queries in descending order by their aggregated mutual information scores. In this work, we select 20 feature terms for each query keyword and then generate all the possible search intentions, from which we further identify the top $k$ qualified and diversified queries w.r.t. the original query. 
 
  
     
  \subsection{Baseline Solution}

 \begin{algorithm}[t]
  \caption{Baseline Algorithm} \label{algo:kqdiversify} 
  \textbf{input:} a query $q$ with $n$ keywords and XML data $T$\\
  \textbf{output:} Top-$k$ search intentions $Q$ and overall result set $\Phi$
    
    \begin{algorithmic}[1]
     \STATE  $M_{m\times n}$ = getFeatureTerms($q$, $T$);\label{line:matrix}
     \WHILE{($q_{new}$ = GenerateNewQuery($M_{m\times n}$)) $\neq$ null} \label{line:gennewquery}
     
     \STATE $\phi$ = null and $prob\_s\_k$ = 1;\label{line:prepare1} 
     
     \STATE $l_{i_xj_y}$ = getNodeList($s_{i_xj_y}$, $T$) for $s_{i_xj_y} \in q_{new} \wedge 1 \leq i_x \leq m  \wedge 1 \leq j_y \leq n$;
     \STATE $prob\_s\_k$ = $\prod_{f_{i_xj_y} \in s_{i_xj_y} \in q_{new}} (\frac{|l_{i_xj_y}|}{getNodeSize(f_{i_xj_y}, T)})$; \label{line:prepare2}
     
     \STATE $\phi$ = ComputeSLCA(\{$l_{i_xj_y}\}$);
     \STATE $prob\_q\_new$ = $prob\_s\_k$ * $|\phi|$; \label{line:probsk}
     
      \IF{$\Phi$ is empty}\label{line:baselinecompare1}
     		\STATE $score(q_{new})$ = $prob\_q\_new$;
     \ELSE
     		
     		\FORALL{Result candidates $r_x \in \phi$}
     				\FORALL{Result candidates $r_y \in \Phi$}
     						\IF{$r_x == r_y$ or $r_x$ is an ancestor of $r_y$}
     							\STATE $\phi.remove(r_x)$;
     						\ELSIF{$r_x$ is a descendant of $r_y$}
     							\STATE $\Phi.remove(r_y)$;
     						\ENDIF     						
     				
     				\ENDFOR
     		\ENDFOR \label{line:baselinecompare2}

     		\STATE $score(q_{new})$ = $prob\_q\_new$ * $|\phi|$* $\frac{|\phi|}{|\phi|+|\Phi|}$;\label{line:finalscore}
     \ENDIF
     
     \IF{$|Q| < k$}\label{line:comparenewquery1}
     			 \STATE put $q_{new}:score(q_{new})$ into $Q$;
     			 \STATE put $q_{new}:\phi$ into $\Phi$; 
     \ELSIF{$score(q_{new}) > score(\{q'_{new}\in Q\})$}  
     				\STATE replace $q'_{new}: score(q'_{new})$ with $q_{new}:score(q_{new})$; 
     				\STATE $\Phi.remove(q'_{new})$;
     \ENDIF \label{line:comparenewquery2}

     	\ENDWHILE
     \RETURN $Q$ and result set $\Phi$;
  \end{algorithmic}
  \end{algorithm} 
  
Given a keyword query, the intuitive idea of baseline algorithm is that we first retrieve the pre-computed feature terms of the given keyword query from the XML data $T$; and then we generate all the possible intended queries based on the retrieved feature terms; at last, we compute the SLCAs as keyword search results for each query and measure its diversification score. As such, the top-$k$ diversified queries and their corresponding results can be returned to users.

Different from traditional XML keyword search, we have to detect and remove the duplicated or ancestor 
results by comparing the new generated results with the previously generated ones. This is because a result may cover multiple search intentions. To meet the requirement of keyword search diversification, however, we are required to return the distinct SLCA results to the users.

The detailed procedure has been shown in Algorithm~\ref{algo:kqdiversify}.  
Given a keyword query $q$ with $n$ keywords, we first load its pre-computed relevant feature terms from the XML data $T$, which is used to construct a matrix $M_{m\times n}$ as shown in Line~\ref{line:matrix}.
 And then, we generate a new query $q_{new}$ from the matrix $M_{m\times n}$ by calling the function GenerateNewQuery() as shown in Line~\ref{line:gennewquery}. The generation of new queries are in the descending order of their mutual information scores. 
 Line~\ref{line:prepare1}-Line~\ref{line:probsk} show the procedure of computing $Prob(q|q_{new}, T)$. 
 To compute the SLCA results of $q_{new}$, we need to retrieve the precomputed node lists of the keyword-feature term pairs in $q_{new}$ from $T$ by $getNodeList(s_{i_xj_y}, T)$.
 Based on the retrieved node lists, we can compute the likelihood of generating the observed query $q$ while the intended query is actually $q_{new}$, i.e., $Prob(q|q_{new},T)$ = $\prod_{f_{i_xj_y}\in s_{i_xj_y} \in q_{new}}$ $(|l_{i_xj_y}|$ $/getNodeSize$ $(f_{i_xj_y},$ $T))$ using Equation~\ref{equ:keywordintent} where $getNodeSize(f_{i_xj_y}, T)$ can be quickly obtained based on the precomputed statistic information of $T$.   
 After that, we can call for the function ComputeSLCA() that can be implemented using any existing XML keyword search method.
In Line~\ref{line:baselinecompare1} - Line~\ref{line:baselinecompare2}, we compare the SLCA results of the current query and the previous queries in order to obtain the distinct and diversified SLCA results. 
At Line~\ref{line:finalscore}, we compute the final score of $q_{new}$ as a diversified  query w.r.t. the previously generated queries in $Q$. 
At last, we compare the new query and the previously generated queries and replace the unqualified ones in $Q$, which is shown in Line~\ref{line:comparenewquery1} - Line~\ref{line:comparenewquery2}.
After processing all the possible queries
, we can return the top $k$ generated queries with their SLCA results.

In the worst case, all the possibe queries in the matrix have the possibility of being chosen as the top-k qualified query candidates. In this worst case, the complexity of the algorithm is O($m^{|q|}*L_1 \sum_{i=2}^{|q|}logL_i$) where $L_1$ is the shortest node list of any generated query, $|q|$ is the number of original query keywords and $m$ is the size of selected features for each query keyword. In practice, the complexity of the algorithm can be reduced by reducing the number $m$ of feature terms, which can be used to bound the number (i.e., reducing the value of $m^{|q|}$) of generated queries.

\subsection{Anchor-based Pruning Solution}   
  
By analysing the baseline solution, we can find that the main cost of this solution is spent on computing SLCA results and removing unqualified SLCA results from the newly and previously generated result sets. 
To reduce the computational cost, we are motivated to design an anchor-based pruning solution, which can avoid the unnecessary computational cost of unqualified SLCA results (i.e., duplicates and ancestors). 
In this subsection, we first analyze the interrelationships between the intermediate SLCA candidates that have been already computed for the generated queries $Q$ and the nodes that will be merged for answering the newly generated query $q_{new}$. And then, we will propose the detailed description and algorithm of the anchor-based pruning solution.

 \subsubsection{Properties of Computing diversified SLCAs} 
  
  

  \begin{definition}\emph{(Anchor nodes)}
  Given a set of queries $Q$ that have been already processed and a new query $q_{new}$, the generated SLCA results $\Phi$ of $Q$ can be taken as the anchors for efficiently computing the SLCA results of $q_{new}$ by partitioning the keyword nodes of $q_{new}$. 
  \end{definition}

    \begin{figure}[htbp]
  \centering   
    \includegraphics[scale=0.5]{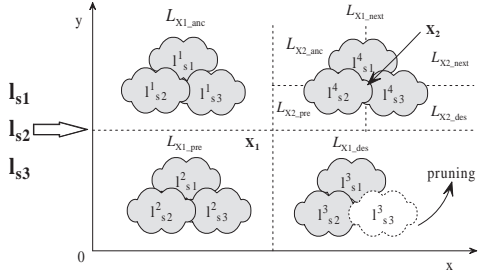}      
  \caption{The usability of anchor nodes}\label{fig:anchor}
	\end{figure} 
  
  
  \begin{example}
  Figure~\ref{fig:anchor} shows the usability of anchor nodes for computing SLCA results.
    Consider two SLCA results $X_1$ and $X_2$ (assume $X_1$ precedes $X_2$) for the current query set $Q$.  
  For the next query $q_{new} = \{s_1, s_2, s_3\}$ and its keyword instance lists $L$ = $\{l_{s1},$ $l_{s2},$ $l_{s3}\}$, the keyword instances in $L$ will be divided into four areas by the anchor $X_1$: (1) $L_{X_1\_anc}$ in which all the keyword nodes are the ancestor of $X_1$ so $L_{X_1\_anc}$ cannot generate new and distinct SLCA results; (2) $L_{X_1\_pre}$ in which all the keyword nodes are the previous siblings of $X_1$ so we may generate new SLCA results if the results are still bounded in the area; (3) $L_{X_1\_des}$ in which all the keyword nodes are the descendant of $X_1$ so it may produce new SLCA results that will replace $X_1$; and (4) $L_{X_1\_next}$ in which all the keyword nodes are the next siblings of $X_1$ so it may produce new results, but it may be further divided by the anchor $X_2$. 
  If there is no intersection of all keyword node lists in an area, then the nodes in this area can be pruned directly, e.g., $l^3_{s1}$ and $l^3_{s2}$ can be pruned without computation if $l^3_{s3}$ is empty in $L_{x1\_des}$. 
  Similarly, we can process $L_{X_2\_pre}$, $L_{X_2\_des}$ and $L_{X_2\_next}$. After that, a set of new and distinct SLCA results can be obtained with regards to the new query set $Q\bigcup q_{new}$.
    \end{example}

  \begin{theorem}\emph{(Single Anchor)}\label{theorem:singleanchor} Given an anchor node $v_a$ and a new query $q_{new}=\{s_1, s_2, ..., s_n\}$, its keyword node lists $L=\{l_{s_1}, l_{s_2}, ...,$ $l_{s_n}\}$ can be divided into four areas to be anchored by $v_a$, i.e., the keyword nodes that are the ancestors of $v_a$, denoted as $L_{v_a\_anc}$; the keyword nodes that are the previous siblings of $v_a$, denoted as $L_{v_a\_pre}$; the keyword nodes that are the descendants of $v_a$, denoted as $L_{v_a\_des}$; and the keyword nodes that are the next siblings of $v_a$, denoted as $L_{v_a\_next}$. We have that $L_{v_a\_anc}$ does not generate any new result; each of the other three areas may generate new and distinct SLCA results individually; no new and distinct SLCA results can be generated accross the areas. 
  
 
  \end{theorem}
  
 For the nodes in the ancestor area, all of them are the ancestors of the anchor node $v_a$. It says that the SLCA candidates they can produce are also the ancestors of $v_a$. Therefore, the area cannot produce any new result due to the existence of $v_a$.
 For the nodes in $L_{v_a\_pre}$ (or $L_{v_a\_next}$), if they can produce SLCA candidates that are bounded in the left-bottom (or right-top) area of $v_a$, then these  candidates are the new SLCA results. This is because there are no ancestor-descendant relationship between the nodes in the area and $v_a$.
For the nodes in $L_{v_a\_des}$, if they can produce SLCA candidates that are bounded in the right-bottom area of $v_a$, then these candidates will be taken as the new and distinct results while $v_a$ will be removed from result set. 

It is obvious that the nodes in the ancestor area does not take part in the generation of SLCA results with the nodes in any other area. Now let's show that the keyword nodes coming from two of the other areas cannot produce new and distinct SLCA results. If we can select some keyword instances from $L_{v_a\_pre}$ and some keyword instances from $L_{v_a\_des}$, then their corresponding SLCA candidate must be the ancestor of $v_a$, which cannot become the new and distinct SLCA result due to the existence of $v_a$ in result set. Similarly, no result can be generated if we select some keyword instances from the other two areas $L_{v_a\_pre}$ and $L_{v_a\_next}$ (or $L_{v_a\_des}$ and $L_{v_a\_next}$).
  
  
  \begin{theorem}\emph{(Multiple Anchors)}\label{theorem:multiple} Given multiple anchor nodes $V_a$ and a new query $q_{new}=\{s_1, s_2, ..., s_n\}$, its keyword node lists $L=\{l_{s_1}, l_{s_2}, ...,$ $l_{s_n}\}$ can be maximally divided into $(3*|V_a|+1)$ areas to be anchored by the nodes in $V_a$. Only the nodes in $(2*|V_a|+1)$ areas can generate new SLCA candidates individually and the nodes in the $(2*|V_a|+1)$ areas are independent to compute SLCA candidates.   
  \end{theorem}
 
 Let $V_a$ = ($x_1$, ..., $x_i$, ..., $x_j$, ..., $x_{|V_a|}$), where $x_i$ precedes $x_j$ for $i < j$. We first partition the space into four areas using $x_1$, i.e., $L_{x_1\_anc}$, $L_{x_1\_pre}$, $L_{x_1\_des}$ and $L_{x_1\_next}$. For $L_{x_1\_next}$, we partition it further using $x_2$ into four areas. We repeat this process until we partition the area $x_{(|V_a|-1)\_next}$ into four areas $x_{|V_a|\_anc}$, $L_{x_{|V_a|}\_pre}$, $L_{x_{|V_a|}\_des}$ and $x_{|v_a|\_next}$ by using $x_{|V_a|}$. Obviously the total number of partitioned areas is $3*|V_a|+1$. As the ancestor areas do not contribute to SLCA results, we end up with $2*|V_a|+1$ areas that need to be processed.
 
Now consider any two different areas $x_{i\_r_1}$ and $x_{j\_r_2}$ where $r_1$ and $r_2$ could be either ``pre'' or ``des'' and $i \leq j$. If $j = |V_a|$, $r_2$ may also be ``next''. If $i = j$, then we know $r_1 \neq r_2$. By Theorem~\ref{theorem:singleanchor}, we know $x_{i\_r_{1}}$ and $x_{j\_r_2}$ cannot produce new and distinct SLCA results; 
If $i < j$, i.e., $x_i$ precedes $x_j$, then we know $x_{j\_r_2}$ is a sub area of $x_{i\_next}$. By Theorem~\ref{theorem:singleanchor} again, we can get that $x_{i\_r_1}$ and $x_{j\_r_2}$ cannot produce new and distinct SLCA results.

\begin{property}
Consider the $(2*|V_a|+1)$ effective areas to be anchored by the nodes in $V_a$. If $\exists s_i \in q_{new}$ and none of its instances appear in an area, then the area can be pruned because it cannot generate SLCA candidates of $q_{new}$. 
\end{property}
  
  The reason is that any area that can possibly generate new results should contain at least one keyword matched instance for each keyword in the query based on the SLCA semantics. Therefore, if an area contains no keyword instance, then the area can be removed definitely without any computation.

 \subsubsection{The Anchor-based Pruning Algorithm}

  \begin{algorithm}[t]
  \caption{Anchor-based Pruning Algorithm} \label{algo:anchorkqdiversify} 
   \textbf{input:} a query $q$ with $n$ keywords and XML data $T$ \\
  \textbf{output:} Top-$k$ query intentions $Q$ and result set $\Phi$
    
    \begin{algorithmic}[1]
     \STATE  $M_{m\times n}$ = getFeatureTerms($q$, $T$);
     \WHILE{$q_{new}$ = GenerateNewQuery($M_{m\times n}$) $\neq$ null}
     \STATE Line~\ref{line:prepare1}-Line~\ref{line:prepare2} in Algorithm~\ref{algo:kqdiversify};
     
            
     \IF{$\Phi$ is not empty} \label{line:anchorheuristic1}
     		\FORALL{$v_{anchor} \in \Phi$} \label{line:anchor1}
     			\STATE get $l_{i_xj_y\_pre}$, $l_{i_xj_y\_des}$, and $l_{i_xj_y\_next}$ by calling for Partition($l_{i_xj_y}$, $v_{anchor}$);
     		 		
     		 		\IF{$\forall l_{i_xj_y\_pre} \neq $ null}
     		  \STATE $\phi'$ = ComputeSLCA($\{l_{i_xj_y\_pre}\}$, $v_{anchor}$);
     		  \ENDIF
     		  	\IF{$\forall l_{i_xj_y\_des} \neq $ null}
     		   \STATE $\phi''$ = ComputeSLCA($\{l_{i_xj_y\_des}\}$, $v_{anchor}$);
     		  \ENDIF
     		   \STATE $\phi$ += $\phi'$ + $\phi''$;

     		   \IF{$\phi''\neq$ null}
     		   		\STATE $\Phi.remove(v_{anchor})$;
     		   \ENDIF 
     			 
     			 \IF{$\exists l_{i_xj_y\_next} =$null}
     			 		\STATE Break the FOR-Loop;
     			 		
     			 \ENDIF
     			 
     			 \STATE $l_{i_xj_y}$ = $l_{i_xj_y\_next}$ for $1 \leq i_x \leq m  \wedge 1 \leq j_y \leq n$;
     		\ENDFOR \label{line:anchor2}
     		
     		
     \ELSE
     
     		\STATE $\phi$ = ComputeSLCA($\{l_{i_xj_y}\}$); \label{line:anchorempty}
     \ENDIF \label{line:anchorheuristic2}
         		
     \STATE $score(q_{new})$ = $prob\_q\_new$ * $|\phi|$* $\frac{|\phi|}{|\Phi| + |\phi|}$;\label{line:repeat21}
     
     \STATE Line~\ref{line:comparenewquery1}-Line~\ref{line:comparenewquery2} in Algorithm~\ref{algo:kqdiversify};
     \ENDWHILE
     \RETURN $Q$ and result set $\Phi$;
  \end{algorithmic}
  \end{algorithm}

Motivated by the properties of computing diversified SLCAs, we design the anchor-based pruning algorithm. The basic idea is described as follows. We generate the first new query and compute its corresponding SLCA candidates as a start point. When the next new query is generated, we can use the intermediate results of the previously generated queries to prune the unnecessary nodes according to the above theorems and property. By doing this, we only generate the distinct SLCA candidates every time. That is to say, unlike the baseline algorithm, the diversified results can be computed directly without further comparison. 
  
  The detailed procedure is shown in Algorithm~\ref{algo:anchorkqdiversify}.
  Similar to the baseline algorithm, we need to construct the matrix of feature terms, retrieve their conrresponding node lists where the node lists can be maintained using R-tree index. And then, we can calculate the likelihood of generating the observed query $q$ when the issued query is $q_{new}$.
  Different from the baseline algorithm, we utilize the intermediate SLCA results of previously generated queries as the anchors to efficiently compute the new SLCA results for the following queries.
  For the first generated query, we can compute the SLCA results using any existing XML keyword search
method as the baseline algorithm does, shown in Line~\ref{line:anchorempty}. Here, we use stack-based method to implement the function ComputeSLCA().

 The results of the first query will be taken as anchors to prune the node lists of the next query for reducing its evaluation cost in Line~\ref{line:anchor1} - Line~\ref{line:anchor2}.
   Given an anchor node $v_{anchor}$, for each node list $l_{i_xj_y}$ of a query keyword  in the current new query, 
we may get three effective node lists $l_{i_xj_y\_pre}$, $l_{i_xj_y\_des}$ and $l_{i_xj_y\_next}$ using R-tree index by calling for the function Partition(). 
If a node list is empty, e.g., $l_{i_xj_y\_pre} = $null, then we don't need to get the node lists for the other query keywords in the same area, e.g., in the left-bottom area of $v_{anchor}$. This is because it cannot produce any SLCA candidates at all. Consequently, the nodes in this area cannot generate new and distinct SLCA results. 
If all the node lists have at least one node in the same area, then we compute the SLCA results by the function ComputeSLCA(), e.g., ComputeSLCA($\{l_{i_xj_y\_des}\}$, $v_{anchor}$) that merges the nodes in $\{l_{i_xj_y\_des}\}$.
If the SLCA results are the descendant of $v_{anchor}$, then they will be recorded as new distinct results and $v_{anchor}$ will be removed from the temporary result set. 
Through Line~\ref{line:repeat21}, we can obtain the final score of the current query  without comparing the SLCA results of $\phi$ with that of $\Phi$.
At last, we need to record the score and the results of the new query into $Q$ and $\Phi$, respectively.
After all the necessary queries are computed, the top-$k$ diversified queries and their results will be returned.
  \subsection{Anchor-based Parallel Sharing Solution}
Although the anchor-based pruning algorithm can avoid unnecessary computation cost of the baseline algorithm, it can be further improved by exploiting the parallelism of keyword search diversification and reducing the repeated scanning of the same node lists.

\subsubsection{Observations}

According to the semantics of keyword search diversification, only the distinct SLCA results need to be returned to users. We have the following two observations.



\textbf{Observation 1:} 
Anchored by the SLCA result set $V_a$ of previously processed queries $Q$, the keyword nodes of the next query $q_{new}$ can be classified into $2*|V_a|+1$ areas. According to Theorem~\ref{theorem:multiple}, no new and distinct SLCA results can be generated across areas. Therefore, the $2*|V_a|+1$ areas of nodes can be processed independently, i.e., we can compute the SLCA results area by area. It can make the parallel keyword search diversification efficient without communication cost among processors.    


\textbf{Observation 2:} 
Because there are term overlaps between the generated queries, the intermediate partial results of the previously processed queries may be reused for evaluating the following queries, by which the repeated computations can be avoided.


 \subsubsection{Anchor-based Parallel Sharing Algorithm}

To make the parallel computing efficiently, we utilize the SLCA results of previous queries as the anchors to partition the node lists that needs to be computed.
By assigning areas to processors, no communication cost among the processors is needed. Our proposed algorithm guarantees that the results generated by each processor are the SLCA results of the current query. In addition, we also take into account the shared partial matches among the generated queries, by which we can further improve the efficiency of the algorithm.

Different from the above two proposed algorithms, we first generate and analyse all the possible queries $Q_{new}$. Here, we use a vector $V$ to maintain the shared query segments among the generated queries in $Q_{new}$. And then, we begin to process the first query like the above two algorithms. When the next query is coming, we will check its shared query segments and explore parallel computing to evaluate the query. 

To do this, we first check if the new query $q_{new}$ contains shared parts $\psi$ in $V$. For each shared part in $\psi$, we need to check its processing status. If the status has already been set as ``processed'', then it says that the partial matches of the shared part have been computed before. 
In this condition, we only need to retrieve the partial matches from previously cached results. Otherwise, it says that we haven't computed the partial matches of the shared part before. We have to compute the partial matches from the original node lists of the shared segments. After that, the processing status will be set as ``processed''.
And then, the node lists of the rest query segments will be processed.
In this algorithm, we also explore parallel computing to improve the performance of query evaluation. At the beginning, we specify the maximal number of processors and the current processor's id (denoted as PID). And then, we distribute the nodes that need to be computed to the processor with PID in a round.
When all the required nodes are reached at a processor, the processor will be activated to compute the SLCA results of $q_{new}$ or the partial matches for the shared segments in $q_{new}$. 
After all active processors complete their SLCA computations, we get the final SLCA results of the query $q_{new}$.  
At last, we can calculate the score of the query $q_{new}$ and compare its score with the previous ones. If its score is larger than that of one of the queries in $Q$, then  the query $q_{new}$, its score $score(q_{new})$ and its SLCA results will be recorded.
After all the necessary queries are processed, we can return the top-$k$ qualified and diversified queries and their corresponding SLCA results. 
The detailed algorithm is not provided due to the limited space.


\section{Experiments}\label{sec:experiments}

In this section, we show the extensive experimental results for evaluating the performance of our baseline algorithm (denoted as \textit{b}aseline \textit{e}valuation \textit{BE}) and anchor-based algorithm (denoted as \textit{a}nchor-based \textit{e}valuation \textit{AE}), which were implemented in Java and run on a 3.0GHz Intel Pentium 4 machine with 2GB RAM running Windows XP. For our anchor-based parallel sharing algorithm (denoted as \textit{ASPE}), it was implemented using six computers, which can serve as six processors for parallel computation.

\subsection{Dataset and Queries}

We use a real dataset, DBLP~\cite{dblp} and a synthetic XML benchmark dataset XMark~\cite{xmark} for testing the proposed XML keyword search diversification model and our designed algorithms. 
The size of DBLP dataset is 971MB and the size of generated XMark dataset is 697MB. Compared with DBLP dataset, the synthetic XMark dataset has varied depths and complex data structures, but as we know, it does not contain clear semantic information due to its synthetic data. Therefore, we only use DBLP dataset to measure the effectiveness of our diversification model in this work. 

For each XML dataset used, we selected some terms based on the following criteria: (1) a selected term should often appear in user-typed keyword queries; (2) a selected term should highlight different semantics when it co-occurs with feature terms in different contexts. Based on the criteria of selection, we chose some terms for each dataset as follows. 
For DBLP dataset, we selected \textit{``database, query, programming, semantic, structure, network, domain, dynamic, parallel''}. And we chose \textit{``brother, gentleman, look, free, king, gender, iron, purpose, honest, sense, metals, petty, shakespeare, weak, opposite''} for XMark dataset. 
By randomly combining the selected terms, we generated a set of keyword queries for each dataset. 
Here we limited the length of each keyword query as two terms because the short queries often contain more ambiguity and can be diversified into more search intentions. 

    
		    

    

\subsection{Efficiency of Diversification Algorithms}

\begin{figure*}[htbp]
  \centering
  \subfigure[q = \{database, network\}]{\label{fig:dbnetwork}
    \includegraphics[scale=0.8]{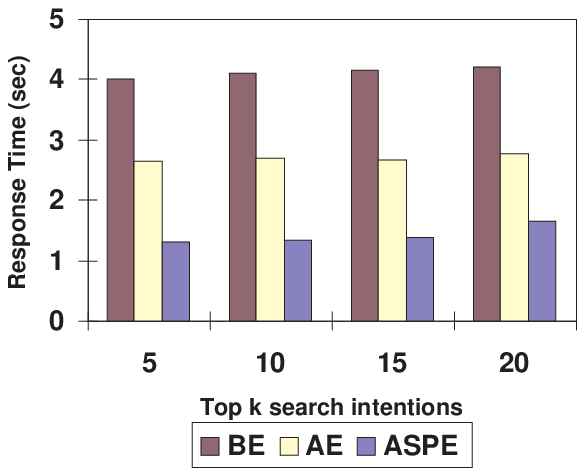}}
  \hskip 0.05in
  \subfigure[q=\{domain, database\}]{\label{fig:domaindb}
    \includegraphics[scale=0.8]{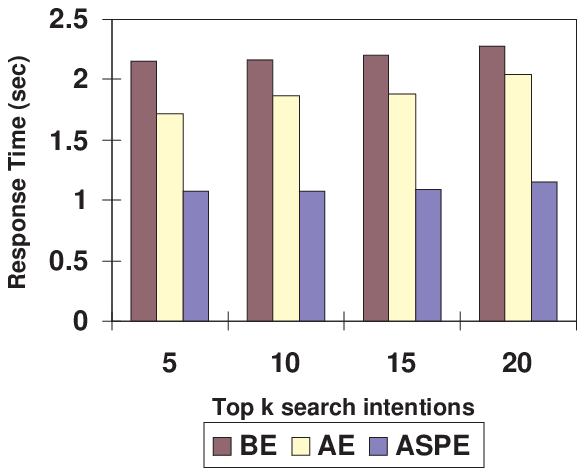}} 
    \hskip 0.05in      
  \subfigure[q=\{domain, query\}]{\label{fig:domainquery}
    \includegraphics[scale=0.8]{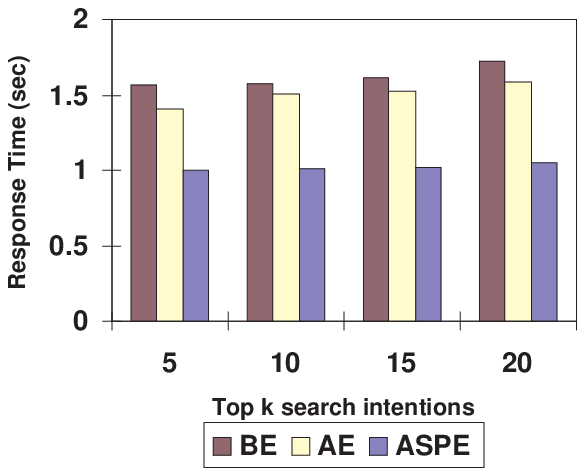}}  
    \\ 
  \subfigure[q=\{dynamic, database\}]{\label{fig:dynamicdb}
    \includegraphics[scale=0.8]{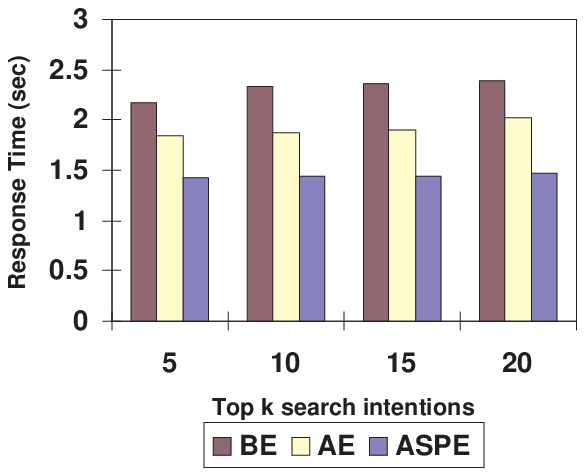}} 
   \hskip 0.05in            
  \subfigure[q=\{dynamic, query\}]{\label{fig:dynamicquery}
    \includegraphics[scale=0.8]{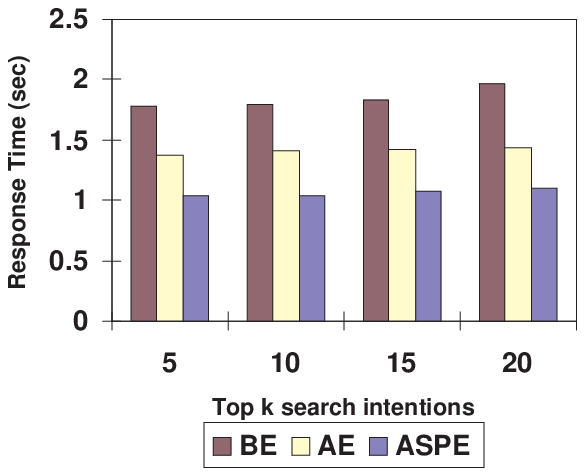}}  
     \hskip 0.05in
  \subfigure[q=\{network, parallel\}]{\label{fig:networkparallel}
    \includegraphics[scale=0.8]{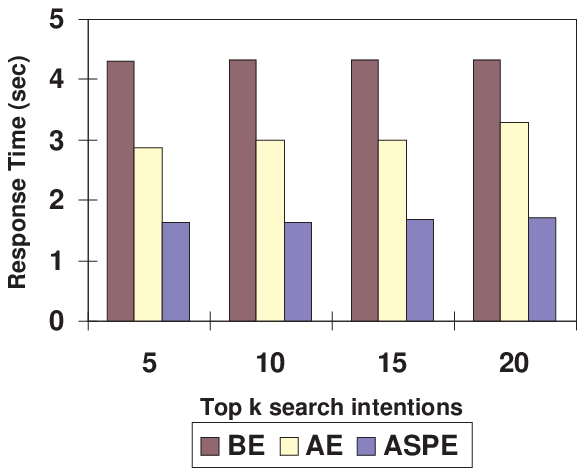}}  
    \\    
  \subfigure[q=\{parallel, database\}]{\label{fig:paralleldb}
    \includegraphics[scale=0.8]{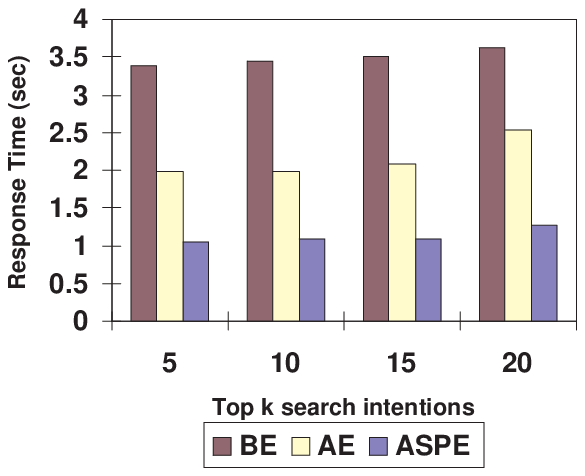}}  
\hskip 0.05in
 \subfigure[q=\{parallel, query\}]{\label{fig:parallelquery}
    \includegraphics[scale=0.8]{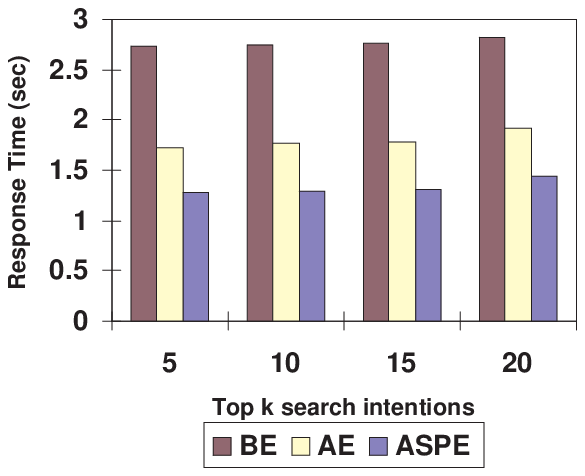}} 
\hskip 0.05in
      \subfigure[q=\{programming, database\}]{\label{fig:programdb}
    \includegraphics[scale=0.8]{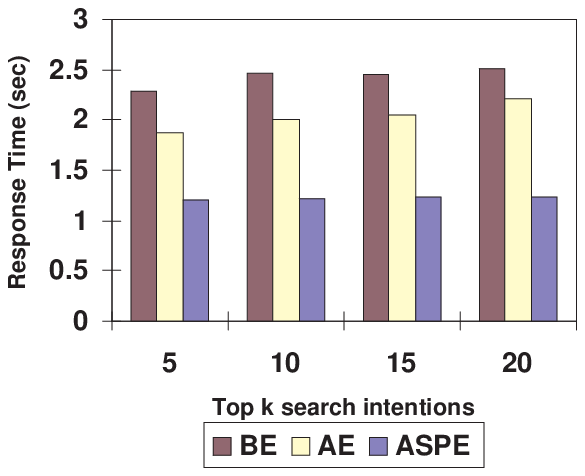}}  
  \\
  \subfigure[q=\{query, database\}]{\label{fig:querydatabase}
    \includegraphics[scale=0.8]{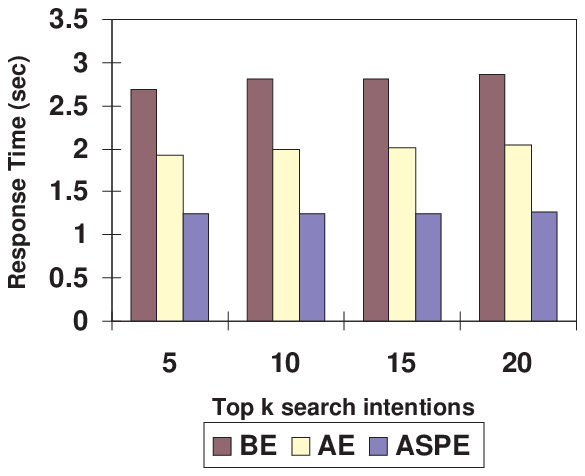}}           
     \hskip 0.05in
  \subfigure[q=\{semantic, query\}]{\label{fig:semanticquery}      
    \includegraphics[scale=0.8]{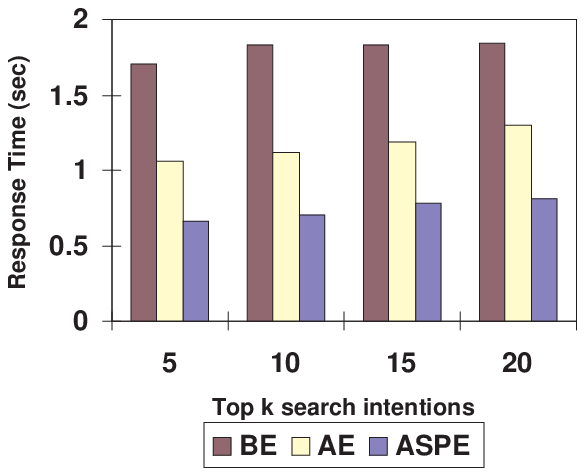}} 
     \hskip 0.05in
       \subfigure[q=\{structure, network\}]{\label{fig:structurenetwork}
    \includegraphics[scale=0.8]{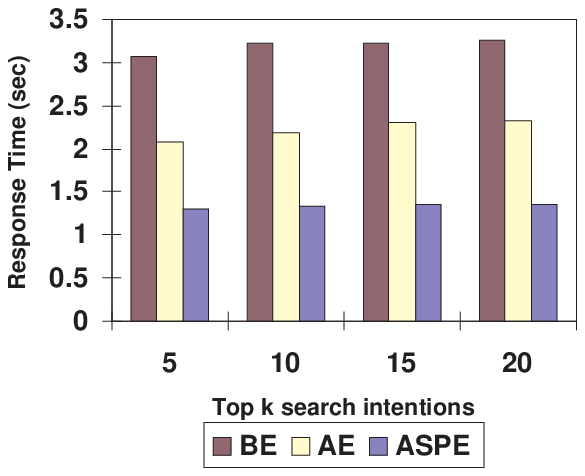}} 
         \caption{12 DBLP queries for k=5, 10, 15, 20}
  \label{fig:varyquery} 
\end{figure*}

\begin{figure*}[htbp]
  \centering
  \subfigure[q = \{brother, gentlemen\}]{\label{fig:brothergent}
    \includegraphics[scale=0.8]{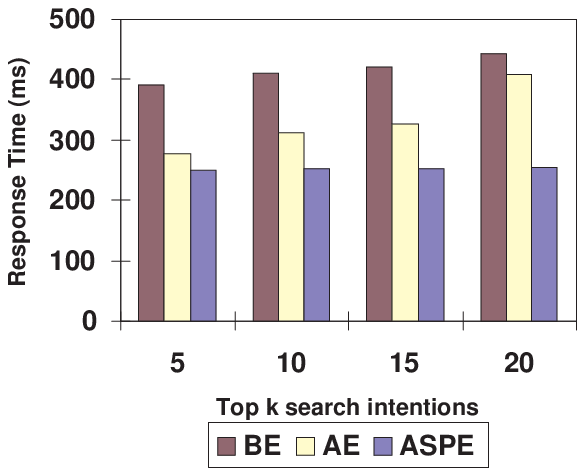}}
  \hskip 0.05in
  \subfigure[q=\{brother, look\}]{\label{fig:brotherlook}
    \includegraphics[scale=0.8]{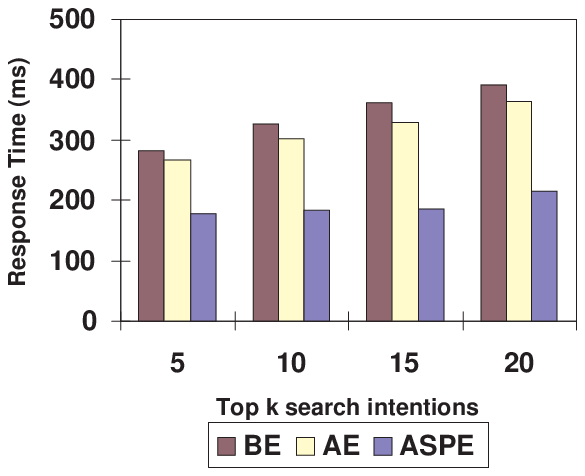}} 
   \hskip 0.05in
  \subfigure[q=\{court, lands\}]{\label{fig:courtlands}
    \includegraphics[scale=0.8]{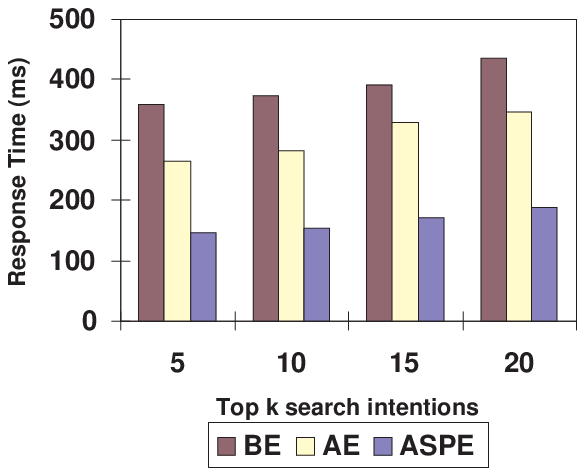}}  
   \\ 
  \subfigure[q=\{free, king\}]{\label{fig:freeking}
    \includegraphics[scale=0.8]{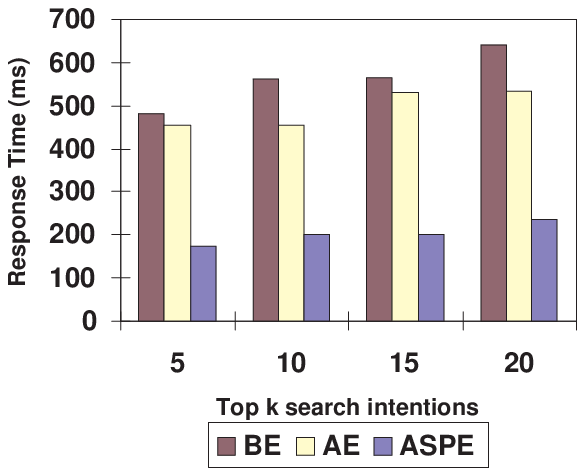}}     
      \hskip 0.05in       
   \subfigure[q=\{shakespeare, king\}]{\label{fig:shakeking}
    \includegraphics[scale=0.8]{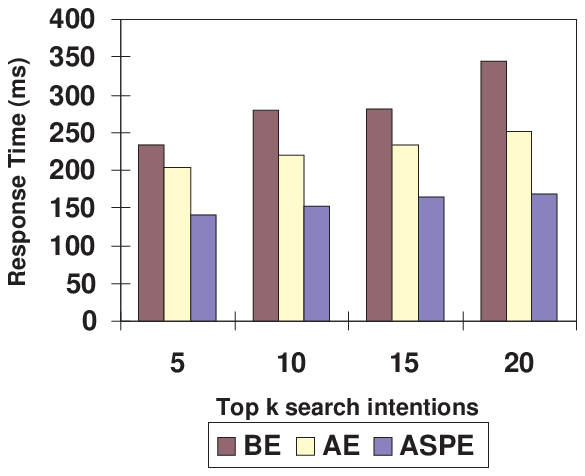}}  
     \hskip 0.05in
       \subfigure[q=\{iron, look\}]{\label{fig:ironlook}
    \includegraphics[scale=0.8]{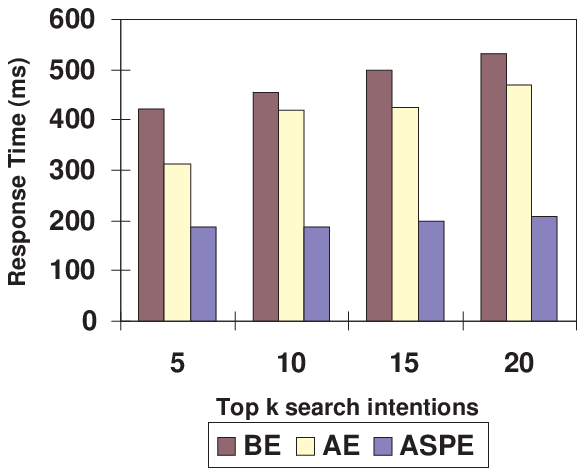}} 
    \\  
  \subfigure[q=\{iron, purpose\}]{\label{fig:ironpurpose}
    \includegraphics[scale=0.8]{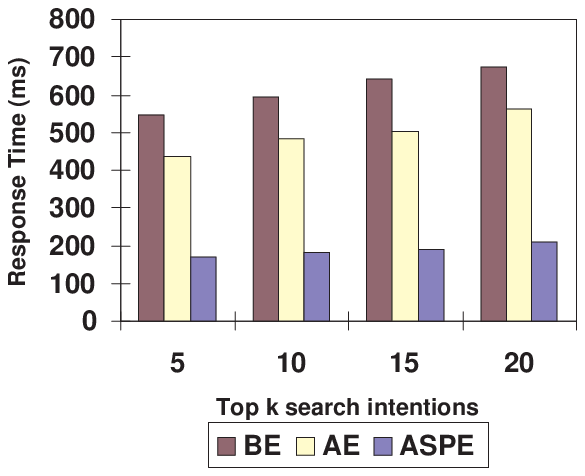}}    
      \hskip 0.05in
  \subfigure[q=\{king, honest\}]{\label{fig:kinghonest}
    \includegraphics[scale=0.8]{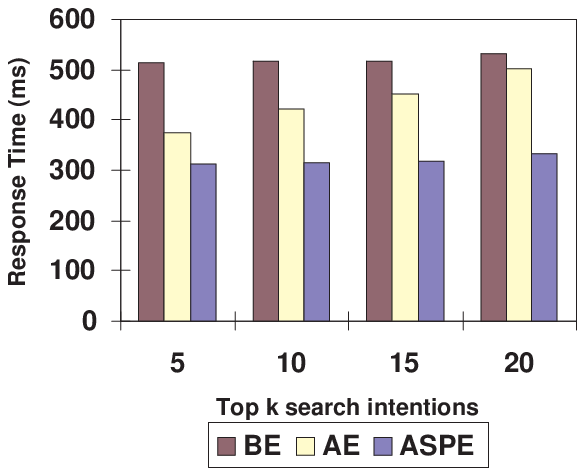}} 
    \hskip 0.05in 
  \subfigure[q=\{live, sense\}]{\label{fig:livesense}
    \includegraphics[scale=0.8]{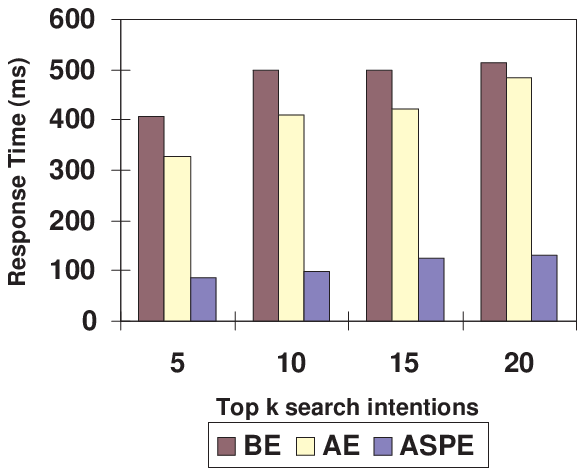}}     
 \\
  \subfigure[q=\{metals, look\}]{\label{fig:metalslook}
    \includegraphics[scale=0.8]{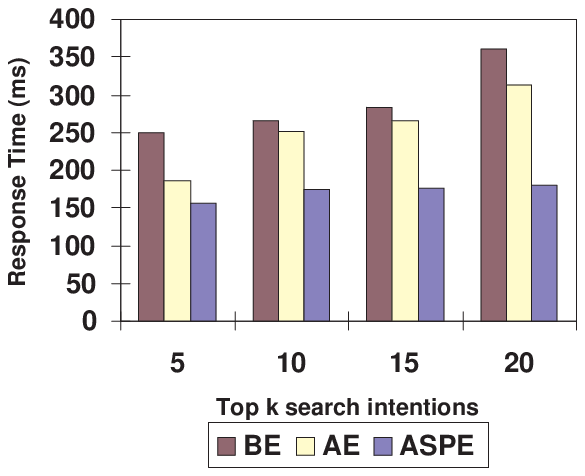}}
     \hskip 0.05in     
    \subfigure[q=\{petty, sense\}]{\label{fig:pettysense}
    \includegraphics[scale=0.8]{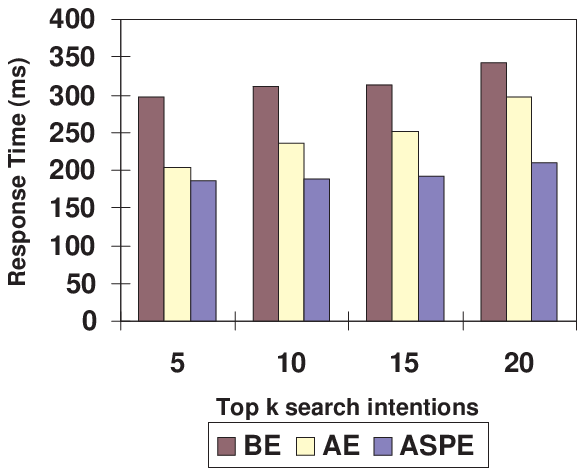}}  
 \hskip 0.05in
 \subfigure[q=\{purpose, look\}]{\label{fig:purposelook}
    \includegraphics[scale=0.8]{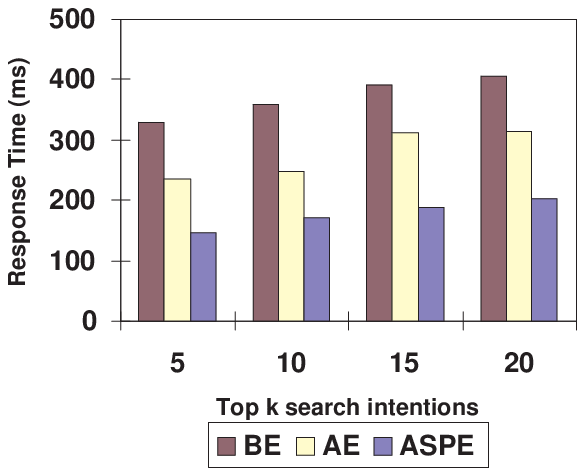}}
 %
         \caption{12 XMARK queries for k=5, 10, 15, 20}
  \label{fig:varyqueryxmark} 
\end{figure*}

We show the efficiency of our proposed diversification algorithms by selecting 12 keyword queries for each dataset due to the limited space. Figure~\ref{fig:varyquery} shows their responce time when we do keyword search diversifiction over DBLP dataset. 
According to the experimental results, BE needs to take about 2.7 seconds to answer an original query on average. However, AE and ASPE can finish in 1.7 seconds and 1.1 seconds on average, respectively.  
Compared with BE, AE can reduce the response time by about 37\% on average and ASPE can reduce the response time by 59.2\% on average. This is because lots of nodes can be pruned without computation. For example, Figure~\ref{fig:paralleldb} shows the response time of the three algorithms for the query \{parallel, database\}, from which we can see that BE takes about 3.5 seconds, AE takes about 2.4 seconds, and ASPE takes about 1.1 seconds. AE outperforms BE because BE needs to process 233,566 nodes while AE only needs to process 177,069 nodes by pruning 56,497 nodes. 
ASPE can further improve the efficiency of keyword search diversification because 
the 177,069 nodes can be processed in parallel. Let's consider another query \{domain, query\} in Figure~\ref{fig:domainquery} where BE and AE consume the similar response time. From the experimental results, we can find that BE processes 114,418 nodes and AE also needs to process 110,687. Here, only 3,731 nodes can be pruned.


  Figure~\ref{fig:varyqueryxmark} shows the experimental results when we do keyword search diversification over XMark dataset. Although the efficiency of BE is slower than that AE and ASPE, it can still finish each query evaluation in 0.7 second. Compared with BE, the improvement of AE is not significant because (1) the size of keyword nodes is not as large as that of DBLP, e.g., most keyword queries can be processed by exploring about 40,000 nodes in XMark dataset while it has to explore about 230,000 nodes to answer most queries in DBLP dataset; (2) the keyword nodes are distributed evenly in the synthetic XMark dataset, which limits the performance of AE and ASPE, e.g., for \{brother, gentlemen\}, BE needs to process 47,000 nodes while AE still needs to deal with 44,541 nodes.

From Figure~\ref{fig:varyquery} and Figure~\ref{fig:varyqueryxmark}, we can find that the increasing number of search intentions just affects the response time of BE, AE and ASPE  a little. This is because for each original keyword query, we first select 20 feature terms and then generate 400 possible search intentions. After that, the generated search intentions have to be evaluated in the descending order of their mutual information scores.

  \begin{figure}[htbp]
  \centering
  \subfigure[DBLP dataset]{\label{fig:dblpaveragetime}
       \includegraphics[scale=0.5]{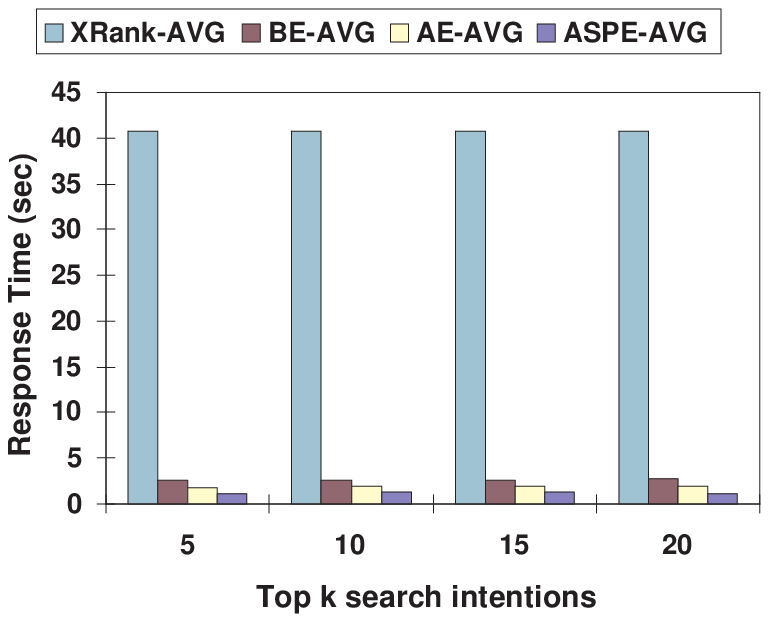}}
  \subfigure[XMARK dataset]{\label{fig:xmarkaveragetime}
       \includegraphics[scale=0.5]{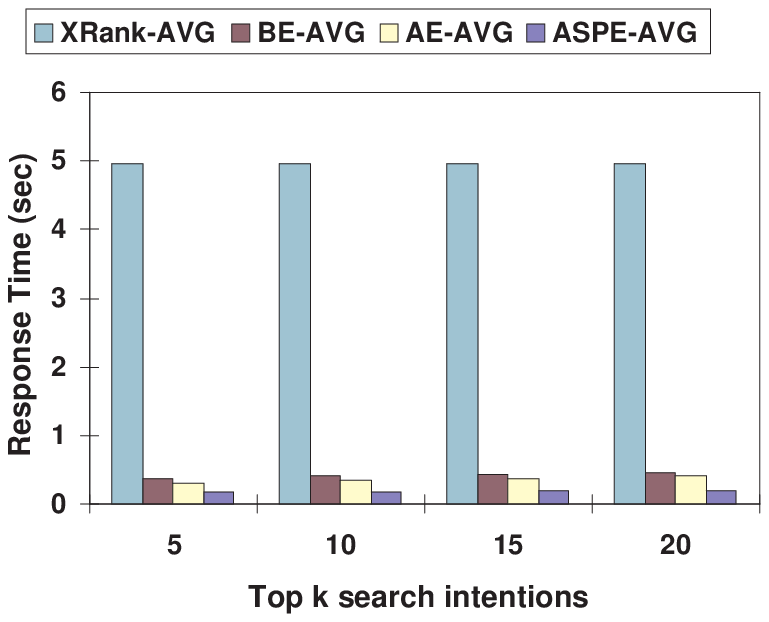}}
                 
  \caption{Average time comparison of original queries and their search intentions}\label{fig:averagetime}
\end{figure} 
 
For showing the experimental performance fairly, we implement the function ComputeSLCA() in BE, AE and ASPE by using the brief approach DIL in XRank~\cite{DBLP:conf/sigmod/GuoSBS03}. Although the other efficient SLCA computation algorithms can be applied to our work, they will require special processing more or less, which is out of our study in this work. 
   Figure~\ref{fig:averagetime} shows the average response time of evaluating the 14 original queries by using XRank~\cite{DBLP:conf/sigmod/GuoSBS03}, and directly computing their diversifications by using our proposed BE, AE and ASPE. According to the experimental results, we can see that if we first compute the keyword search results of the original keyword query and then select the diversified results, it will take at least 10 times of time cost of our proposed diversification approaches. Especially, when the users are only interested in a few of contexts that are distributed in the data to be searched, our context-based keyword query diversification algorithms can outperform the post-processing diversification approaches greatly.

\subsection{Effectiveness of Diversification Model}

\begin{table*}[ht] 
  \renewcommand{\arraystretch}{1.2}
  \small
  \centering
  \caption{Selected top-5 search intentions over DBLP dataset}
  \label{tab:searchintention}
    \scalebox{0.8}{
  \begin{tabular}{|l|l|c|}
      \hline
      {original query} &  {top 5 search intentions} & \% of original query results \\ 
      \hline
       \multirow{2}{*}{	\{semantic query\} }& \{semantic query \underline{reformulation}\}; \{semantic query \underline{languages}\}; \{semantic query \underline{optimization}\}; & \multirow{2}{*}{16.3\%} \\
       	& \{semantic query \underline{processing}\}; \{semantic \underline{models} \underline{relational} database\}& \\
           	 	\hline
           	 	
  \multirow{2}{*}{	\{structure query\}}& \{structure query \underline{performance}\}; \{\underline{spatial} structure query \underline{efficient}\}; \{structure query \underline{language}\}; & \multirow{2}{*}{15.6\%} \\
  	& \{structure query \underline{reformulation}\}; \{structure query \underline{evaluation}\}&\\
  	\hline
  	
     \multirow{2}{*}{ \{domain query\}}& \{domain query \underline{language}\}; \{query \underline{performance} \underline{specific} domain\}; \{\underline{specific} domain query \underline{languages}\}; & \multirow{2}{*}{13\%} \\
      & \{query \underline{expansion} domain \underline{identification}\};  \{query \underline{processing} domain \underline{database}\}& \\ 
  	\hline
  	
  	\multirow{2}{*}{\{dynamic query\}}& \{dynamic query \underline{languages}\}; \{dynamic query \underline{approach}\}; \{query \underline{processing} dynamic \underline{database}\}; & \multirow{2}{*}{12.1\%} \\
  	& \{dynamic query \underline{efficient}\}; \{dynamic query \underline{evaluation}\};&\\
  	\hline 
  	 	
  \multirow{2}{*}{*\{parallel query\}} & \{parallel query \underline{optimization}\}; \{parallel query \underline{language}\}; \{\underline{ranking} query parallel \underline{algorithm}\};& \multirow{2}{*}{11.4\%} \\
  	& \{query \underline{optimization} parallel \underline{database}\}; \{parallel query \underline{performance}\}&\\
  	 \hline
  	 
      \multirow{2}{*}{\{domain database\} }& \{\underline{image} database \underline{specific} domain\}; \{\underline{video} database \underline{specific} domain\}; \{\underline{relational} database \underline{multi} domain\}; & \multirow{2}{*}{10\%} \\
      & \{\underline{protein} database domain\};  \{\underline{object} database domain\}&\\
        	 
  	\hline
       			       
      \multirow{2}{*}{\{database, network\}} & \{\underline{protein} database network\}; \{\underline{mobile} database \underline{wireless} network\}; \{database \underline{management} \underline{sensor} network\}; & \multirow{2}{*}{6.5\%} \\ 
      & \{\underline{distributed} database network\}; \{\underline{relational} database network\}&  \\ 
     	\hline  
     		
  \multirow{2}{*}{\{semantic database\}} & \{semantic \underline{query} database \underline{approach}\}; \{semantic \underline{relationships} \underline{object} database\}; \{\underline{learning} semantic \underline{object} database\};& \multirow{2}{*}{6.3\%} \\
  	& \{semantic \underline{data} \underline{distributed} database\} &\\  	
  	
  	\hline  	
  	\multirow{2}{*}{\{parallel database\} }& \{parallel \underline{processing} \underline{protein} database\}; \{parallel \underline{query} \underline{object} database\}; \{parallel \underline{scheduling} database \underline{query}\};& \multirow{2}{*}{4.2\%} \\
  	& \{parallel \underline{performance} database \underline{applications}\}; \{parallel \underline{ODMG} \underline{object} database\}&\\

      \hline  	
  	\multirow{2}{*}{\{dynamic database\} }& \{dynamic \underline{sequence} database\}; \{dynamic \underline{relational} database\}; \{dynamic \underline{location} database \underline{management}\};& \multirow{2}{*}{4\%} \\
  	& \{dynamic \underline{allocation} \underline{distributed} database\}; \{dynamic database \underline{approach}\}&\\
  	
  		\hline
  	 	
  \multirow{2}{*}{	\{structure network\}}& \{\underline{protein} structure network\}; \{structure \underline{discovery} network \underline{data}\}; \{structure \underline{feature} network \underline{data}\};& \multirow{2}{*}{3.3\%} \\
  	& \{structure \underline{content} \underline{neural} network\}; \{structure network \underline{database}\}& \\
  		\hline
  	
  \multirow{2}{*}{*\{query database\} }& \{query \underline{language} database \underline{applications}\}; \{query \underline{processing} \underline{relational} database\}; \{query \underline{engine} \underline{relational} database\}; & \multirow{2}{*}{2.5\%} \\
  	& \{query \underline{optimization} \underline{object} database\}; \{query \underline{efficient} \underline{relational} database\}& \\
  	    	\hline
  	
  \multirow{3}{*}{*\{programming database\} }&\{\underline{logic} programming database \underline{approach}\};   \{\underline{logic} programming database \underline{applications}\}; & \multirow{3}{*}{1.9\%} \\
  & \{\underline{dynamic} programming \underline{object} database\}; \{\underline{linear} programming \underline{relational} database\};& \\
  & \{\underline{dynamic} programming \underline{relational} database\} & \\

  	\hline
  	\multirow{2}{*}{\{parallel network\}} & \{parallel \underline{performance} \underline{large} network\}; \{parallel \underline{database} \underline{large} network\}; \{parallel \underline{performance} \underline{peer} network\}; & \multirow{2}{*}{1.7\%} \\
  	& \{parallel \underline{processing} \underline{neural} network\}; \{parallel \underline{scheduling} network\} &\\

  
  	\hline

    \end{tabular}}
\end{table*}

 Table~\ref{tab:searchintention} shows the top 5 search intentions when we evaluate the 14 selected keyword queries over DBLP dataset. At the same time, we also show the percentages of the recommended top 5 search intentions recalling for the searched results of their original queries. 
 Although the generated query of each search intention may contain more keywords than its original query, the results of the generated query and its corresponding results of the original query map to the same SLCA nodes in most cases. 
 This is because the DBLP XML tree is not deep. Therefore, it is fair to compare the SLCA results of original queries with those of their generated queries based on search intentions over DBLP dataset.  
Here, we use the average number of results and average selective rates to do analysis. For example, \{query, database\} can produce more than 2000 results but \{dynamic, database\} can only generate 225 results in total. From the experimental results, we can see that the selective rates of top 5 search intentions are no less than 10\% for the first 6 queries, in the range of (5\%-10\%) for the next 2 queries; and lower than 5\% for the rest 6 queries. Based on our diversification model, for each of the first 6 queries, we can return 22 of 171 distinct results to match with the top 5 search intentions on average. Similarly, we can return about 17 of 250 novel results for each of the next 2 queries and about 15 of 480 distinct results for each of the last 6 queries on average. From the analysis, we can find that our diversification model is not biased to the quantity of results, which has a relatively fixed selective rates, i.e., the percentages of the original query results. Based on the selective rates, it becomes more practicable and easier for users to go through the small number of returned dinstinct results, rather than be overwhelmed with the large number of all results.

 Besides the above analysis at high level, now let's take a look of three cases in more details. For \{parallel query\}, the queries of top 5 selected search intentions can return the publications that discuss the research problems in the different contexts: (1) parallel query \underline{optimization} that highlights \textit{query optimization} problem in general parallel systems or algorithms;
 (2) parallel query \underline{language} that highlights the \textit{query language models} in general parallel systems or algorithms;
 (3) \underline{ranking} query parallel \underline{algorithm} that highlights the \textit{ranking query} in \textit{parallel algorithms};
 (4) query \underline{optimization} parallel \underline{database} that focuses on the query optimization problem on parallel databases;
 (5) parallel query \underline{performance} that focuses on the query \textit{performance analysis} in parallel systems or platforms. 
 Although the context of (4) overlaps with that of (1), the search intention in (4) is clearer than (1) because (4) specifies the database area. 
 For the given query \{programming database\}, the search intentions can be identified by highlighting the context of programming, e.g., logic, linear and dynamic, and the context of database, e.g., relational and object. 
For the given query \{query database\}, it highlights the search intentions in the contexts: query language in database application; query system engine over relational database; query processing and efficiency over relational database; and the query optimization techniques in object database. 

 \section{Related Work}\label{sec:relatedwork}

Recently, diversifying results of document retrieval has been introduced \cite{DBLP:conf/sigir/CarbonellG98,DBLP:conf/wsdm/AgrawalGHI09,DBLP:conf/sigir/ChenK06,DBLP:conf/sigir/ClarkeKCVABM08}.
Most of the techniques perform diversification as a post-processing or re-ranking step of
document retrieval. These techniques first retrieve relevant results and then filter
or re-order the result list to achieve diversification.
\cite{DBLP:conf/sigir/CarbonellG98} is a classic example of such a 
strategy, which can be employed to re-rank documents and promote diversity 
based on maximal marginal relevance (MMR). In \cite{DBLP:conf/wsdm/AgrawalGHI09}, the search results are diversified through 
categorization according to the existing taxonomy of information, in 
which user intents are modelled at the topical level of the taxonomy.
In \cite{DBLP:conf/sigir/ChenK06}, Chen and Karger use Bayesian retrieval
models and condition selection of subsequent documents by making assumptions
about the relevance of the previously retrieved documents. While their
approach is capable of selecting anywhere between $0 < k \leq n$ relevant
documents, they focus primarily on optimizing single document (k=1) and 
perfect precision (k=n) scenarios. Their model does not explicitly consider
user intent or document categorizations.
In \cite{DBLP:conf/sigir/ClarkeKCVABM08}, the authors also consider meaningful ways (Classic ranked retrieval metrics) to evaluate the performance of search diversification and subtopic retrieval algorithms,  such as normalized discounted cumulative gain (NDCG), mean 
average precision (MAP), and mean reciprocal rank (MRR) are discussed by taking user intent into account. However, the above work is predicated 
implicitly on the assumption that a single relevant document will fulfill
a user's information need, making them inadequate for many information
queries. To fix the research gap, 
the authors in \cite{DBLP:conf/www/WelchCO11} present a search diversification
algorithm particularly suitable for informational queries by explicitly
modeling that the user may need more than one page to satisfy her/his need.  
However, the majority of the above work focus on the effectiveness of 
diversifying search results. To improve the efficiency of document retrieval 
with the consideration of diversification, the authors in \cite{DBLP:conf/sigmod/AngelK11} propose an efficient algorithm for diversity-aware search based on low-overhead data access prioritization scheme with theoretical quality guarantees.

Another conventional approach to achieve diversification is clustering or 
classification of search results. By grouping search results
based on similarity, users can navigate to the right groups to 
retrieve the desired information.
Clustering and classification have been applied to document retrieval 
\cite{DBLP:conf/wsdm/AgrawalGHI09}, image retrieval \cite{DBLP:conf/www/LeukenPOZ09}, and database query results \cite{DBLP:conf/sigmod/ChenL07,DBLP:journals/pvldb/LiuJ09,DBLP:journals/pvldb/LiuNC11}. Similar to 
result re-ranking, clustering is usually performed as a post-processing
step, and is computationally expensive. In addition, the common approach of
clustering the most relevant documents, and presenting at most one 
document per cluster, corresponds to a specific, and very limited form of diversification semantics. 

Apart from the above approaches, there are another two work for addressing diversification of search results for document retrieval at the experimental level. \cite{DBLP:conf/www/GollapudiS09} formally explores the axioms that any diversification system should be expected to satisfy, which can be taken 
 as some basis of comparison between different objective functions for 
 diversifying searched results.
\cite{DBLP:conf/sigir/WangZ09} focuses on a theoretical development of the 
economic portfolio theory for document ranking. Their model considers a ``risk'' tradeoff between the expected relevance of a set of documents and correlation between them, modeled as the mean and variance. By balancing the overall relevance of the list against its risk (variance), top-$n$ 
documents will be selected and ordered.

\section{Conclusions}\label{sec:conclusion}
In this paper, we first presented an approach to search diversified results of keyword query from XML data based on the contexts of the query keywords in the data. 
The diversification of the contexts were measured by exploring their relevance to the original query and the novelty of their results.
Furthermore, we designed three efficient algorithms based on the observed properties of XML keyword search results. Finally, we demonstrated the efficiency of our proposed algorithms by running substantial number of queries over both DBLP and XMark datasets. Meanwhile, we also verified the effectiveness of our diversification model by analyzing the returned search intentions for the given keyword queries over DBLP dataset. From the experimental results, we get that our proposed diversification algorithms can return qualified search intentions and results to users in a short time.   



\bibliographystyle{IEEEtran}
\bibliography{vldb_diversification}  


\end{document}